\documentclass{article}

\usepackage[final]{neurips_2021}

\usepackage[utf8]{inputenc} % allow utf-8 input
\usepackage[T1]{fontenc}    % use 8-bit T1 fonts
\usepackage{hyperref}       % hyperlinks
\usepackage{url}            % simple URL typesetting
\usepackage{booktabs}       % professional-quality tables
\usepackage{amsfonts}       % blackboard math symbols
\usepackage{nicefrac}       % compact symbols for 1/2, etc.
\usepackage{microtype}      % microtypography
\usepackage{xcolor}         % colors
\usepackage{graphics}
\usepackage{graphicx}
\usepackage{multirow}

\title{Improving mathematical questioning in teacher training}

\author{%
  Debajyoti Datta\\
  SEAS, University of Virginia\\
  Charlottesville, VA 22903 \\
  \texttt{dd3ar@virginia.edu} \\
  \And Maria Phillips \\
   SEAS, University of Virginia\\
   Charlottesville \\
   USA \\
   \And James P Bywater \\
   James Madison University\\
   Harrisonburg, VA 22807 \\
   USA \\
    \And Jennifer Chiu \\
   School of Education and Human Development\\
   University of Virginia\\
   Charlottesville, VA 22903 \\
   USA \\
    \And Ginger S. Watson \\
   School of Education and Human Development\\
   University of Virginia\\
   Charlottesville, VA 22903 \\
   USA \\
    \And Laura E. Barnes \\
  SEAS, University of Virginia\\
  Charlottesville, VA 22903 \\
  USA \\
    \And Donald E Brown \\
  SEAS, University of Virginia\\
  Charlottesville, VA 22903 \\
  USA \\

}

\begin{document}

\maketitle

\begin{abstract}
High-fidelity, AI-based simulated classroom systems enable teachers to rehearse effective teaching strategies. However, dialogue-oriented open-ended conversations such as teaching a student about scale factors can be difficult to model. This paper builds a text-based interactive conversational agent to help teachers practice mathematical questioning skills based on the well-known Instructional Quality Assessment. We take a human-centered approach to designing our system, relying on advances in deep learning, uncertainty quantification, and natural language processing while acknowledging the limitations of conversational agents for specific pedagogical needs. Using experts' input directly during the simulation, we demonstrate how conversation success rate and high user satisfaction can be achieved.
\end{abstract}

\section{Introduction}

\begin{figure}[t]
\centering
\includegraphics[width=1\textwidth]{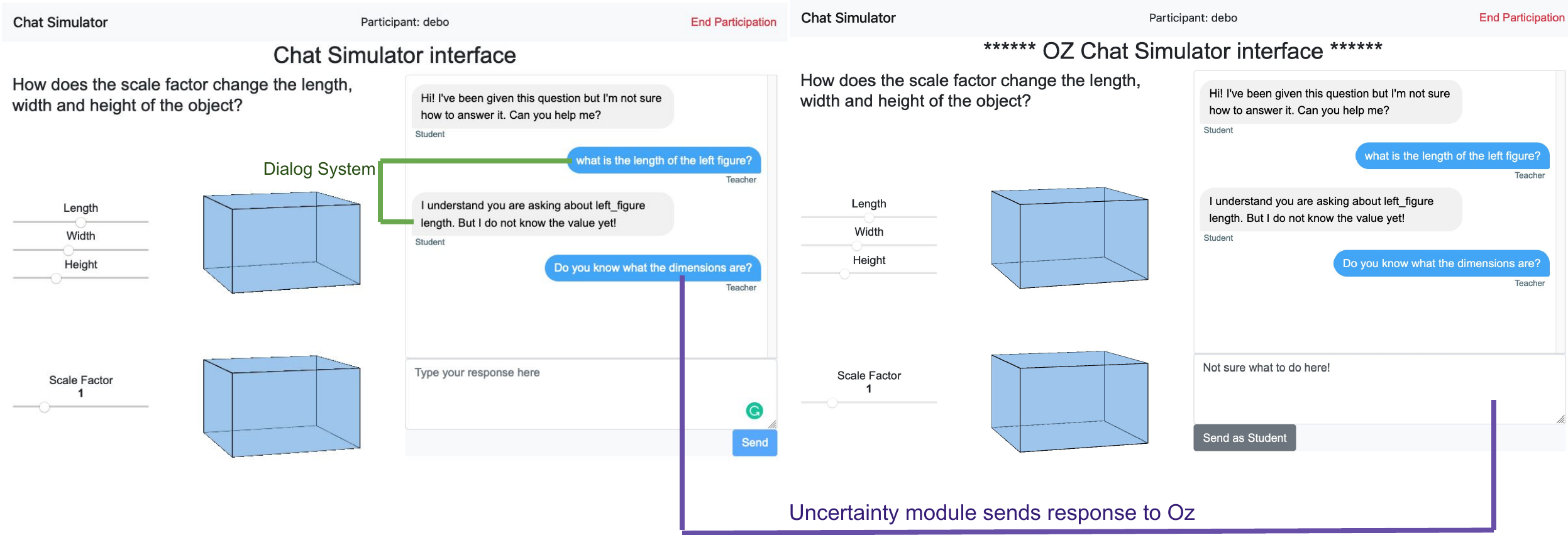} 
% \includegraphics[width=0.3\textwidth]{images/SupervisorRoles.png} 
 % Reduce the figure size so that it is slightly narrower than the column. Don't use precise values for figure width.This setup will avoid overfull boxes.
\caption{\textbf{The user of the system (teacher) and the supervisor have access to a similar interface. The user types an utterance on the left panel, and the dialogue system responds to that utterance. If any dialogue system module fails, the system prompts the supervisor (the right panel) to respond. The supervisor only responds to that specific utterance to prevent the conversation from derailing.}}
\label{fig_scale_factor}
\end{figure}

Significant amounts of improvements in natural language understanding \citep{brown2020language} and common-sense reasoning \citep{sap2020commonsense}, have come from large datasets and deep learning models. However, domain-specific conversational agents still suffer from the lack of large datasets and well-known challenges of deep learning models. One approach the community heavily relies on to mitigate the data scarcity and model fragility problem is to not rely only on AI models but also on human-AI collaborations. Human AI collaborations have been used for AI explanations \cite{bilgic2005explaining, feng2019can, kaur2020interpreting}, content moderation \cite{jhaver2019does} and even dialogue systems \cite{li2016dialogue}. The goal is to achieve complementary performance \cite{bansal2021does}; that is, the system needs to outperform human and AI acting alone. Understanding and improving collaboration between humans and AI systems is tricky because capabilities are hard to calibrate. \cite{nado2021uncertainty}. In this work, we explore Human-AI collaboration to address teacher questioning strategies. The teacher in training uses the system to practice the IQA questioning strategies Table \ref{iqa_description} and the student is a conversational agent. The teacher and the student interact using the text input box as shown in Figure \ref{fig_scale_factor}

% Effective questioning strategies in mathematics classrooms improve student learning outcomes \citep{kilgo2015link, ellis1993teacher, cotton1988classroom, wilen1986effective}, but practicing such questions in a systematic and deliberate way is difficult. For example in a traditional Conversational Agent (CA) for flight booking you might only have to answer very domain specific questions like ``Where are you flying to?'', but in a mathematics classrooms questions are generally more open-ended like ``How did you get that answer''. Similarly, probing questions, as in Table \ref{iqa_description} are often difficult to evaluate because there is a subjective component which depends significantly on the context and the previous utterances.
% In the education domain CAs have resulted in a variety of learning outcomes, \citep{d2014confusion} including improved learning of mathematics concepts like Mathbot \citep{graesser2014learning, grossman2019mathbot} and writing skills \citep{li2021impact}. None of these incorporate a human centered approach for facilitating a task specific conversation.

\begin{figure*}[b]
\centering
\includegraphics[width=0.6\textwidth]{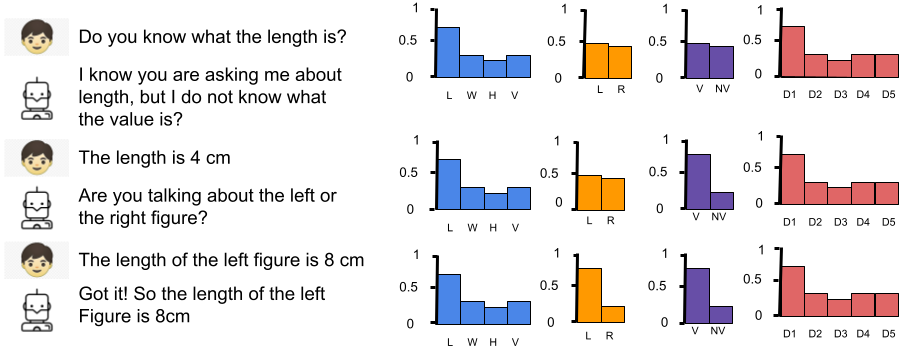} 
\includegraphics[width=0.3\textwidth]{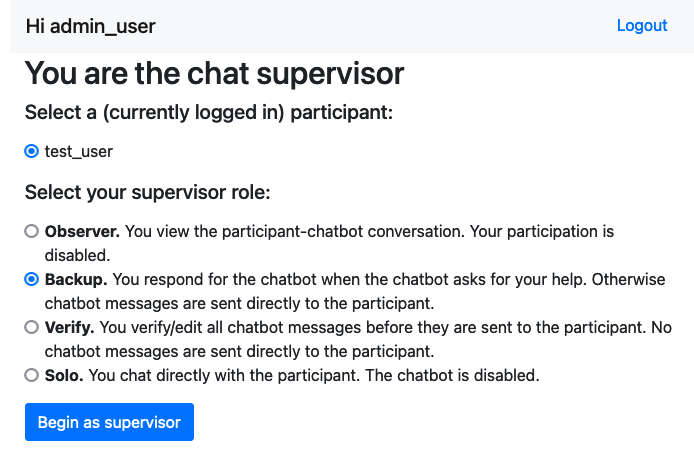} 
 % Reduce the figure size so that it is slightly narrower than the column. Don't use precise values for figure width.This setup will avoid overfull boxes.
\caption{\textbf{The entity recognition stage computes the uncertainty for individual entities. D1-D5 represents dialogue act uncertainty, L,W,H,V represents uncertainty about the dimensions of the object, L, R refer to the uncertainty in figure reference (whether it is a left figure or a right figure), and V, NV refer to whether the attribute mentioned has a value associated with it or not.
}}
\label{fig_uncertainty}
\end{figure*}

\section{Difficulty in making conversational agents in educational settings}

\noindent\textbf{Model and Data Limitations (MDL)}: CA systems that use state of the art deep learning and machine learning also have failure modes. Despite the prevalence of deep learning and new advancements in CA architecture, only recently have limitations been uncovered. Deep learning models (models that are used in most modern CAs) can fail catastrophically because they work by exploiting spurious relationships within data sets which is known as shortcut learning \citep{geirhos_shortcut_2020}. These limitations also exist in Natural Language Processing \citep{poliak2018hypothesis, mccoy2019right}, which makes relying on data and models to build extremely robust systems challenging, especially in domains with relatively less model training data like education. 

\noindent\textbf{User Expectation Limitations (UEL)}: Setting user expectation in task specific CA's is difficult. Norman described the ‘gulfs  of execution and evaluation’ \citep{norman1991cognitive}, as the degree to which the system representations can be perceived and decoded by the user into accurate expectations and intentions of use. Long descriptive answers make designing CAs for teaching scenario difficult as most traditional CA's generate short-answers or factual tasks (``What is 2+2?'').  The ``deep gulf of evaluation'' \citep{norman1991cognitive} exists because CA systems lack meaningful feedback regarding the systems intelligence and capability. This mismatch of user expectations and current technology capabilities can lead to a mis-application of CAs and a lack of confidence that results in an avoidance of use and deployment for complex tasks or sensitive activities.

\begin{table*}[]
\caption{\label{iqa_description}
\textbf{Example teacher utterances and IQA categories. Note that some utterances like ``Sit Down'', or ``Close the book'', or ``Look at the diagram'' are very classroom specific and teachers do not use such utterances while conversing with a CA. }
}
\scalebox{0.7}{

\begin{tabular}{lll}
\hline
\textbf{Question label}                                                                                       & \textbf{Description}                                                                                                                                                                                                                                                                 & \multicolumn{1}{l}{\textbf{Examples}}                                                                                                                                                                                                                                                                      \\ \hline
\begin{tabular}[c]{@{}l@{}}Probing or \\ exploring mathematical\\  meanings and \\ relationships\end{tabular} & \begin{tabular}[c]{@{}l@{}}Question clarifies student thinking, enables \\ students to elaborate their own thinking for \\ their own benefit and the class; Points to \\ underlying mathematical relationships\\  and meanings; Makes links among \\ mathematical ideas\end{tabular} & \multicolumn{1}{l}{\begin{tabular}[c]{@{}l@{}}How did you get that answer?\\ Explain to me how you got that expression?\\ What does n represent in terms of the diagram?\\ Why is it staying the same?\end{tabular}}  \\ \hline
Factual or recall                                                                                             & \begin{tabular}[c]{@{}l@{}}Elicits a mathematical fact; Requires \\ a single response answer; Requires \\ the recall of a memorized fact or \\ procedure, can be a yes/no answer but \\ for a specific mathematical question\end{tabular}                                            & \begin{tabular}[c]{@{}l@{}}What is 3x5?\\ Does this picture show ½ or ¼?\\ What do you subtract first?\end{tabular}                                                                                                                          \\ \hline
Expository or cueing                                                                                          & \begin{tabular}[c]{@{}l@{}}Provides mathematical cueing or \\ mathematical information to students, \\ tells them to look at specific information\\  without engaging students’ ideas\end{tabular}                                                                                   & \begin{tabular}[c]{@{}l@{}}Rhetorical questions (“The answer is three, right?”)\\ Clarifying statements “Between the 2?”\\ Look at this diagram\end{tabular}                                                                                                      \\ \hline
Other                                                                                                         & \begin{tabular}[c]{@{}l@{}}Non-academic behavioral talk; General \\ classroom management; everything else.\end{tabular}                                                                                                                                                              & \begin{tabular}[c]{@{}l@{}}Sit down\\ Close your books\end{tabular}                                                                                                                                                              \\ \hline
\end{tabular}}
\end{table*}

% \begin{table}
% \centering

% \caption{\label{perform_comparison} \textbf{IQA data collection was performed with two annotators using the labelstudio data labeling platform \citep{LABELSTUDIO}}}
% \label{perform_comparison}
% \scalebox{0.7}{
% \begin{tabular}{cccc} 
% \hline
% \textbf{Annotator} & \textbf{n} & \begin{tabular}[c]{@{}c@{}}\textbf{Annotator }\\\textbf{ Accuracy}\end{tabular} & \begin{tabular}[c]{@{}c@{}}\textbf{ Agreement}\\\textbf{ (kappa)}\end{tabular}  \\ 
% \hline
% Both               & 1730       & 0.82                                                                            & \multirow{3}{*}{0.52}                                                           \\ 
% \cline{1-3}
% A1                 & 864        & 0.89                                                                            &                                                                                 \\ 
% \cline{1-3}
% A2                 & 866        & 0.74                                                                            &                                                                                 \\
% \hline
% \end{tabular}}
% \end{table}

\begin{figure*}[t]
\centering
\includegraphics[width=1\textwidth]{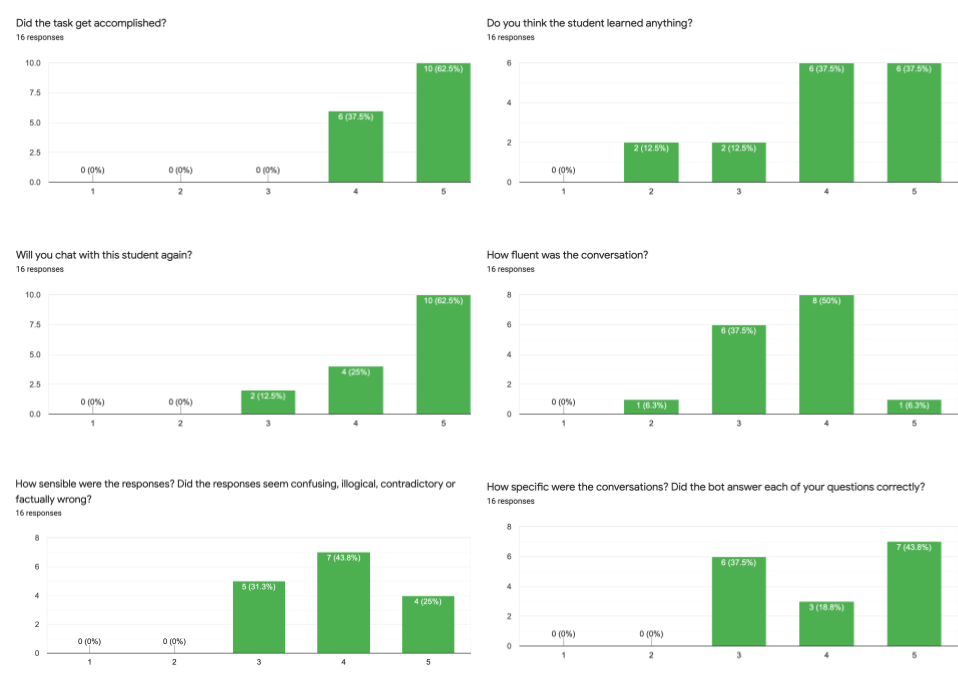} 
 % Reduce the figure size so that it is slightly narrower than the column. Don't use precise values for figure width.This setup will avoid overfull boxes.
\caption{\textbf{Survey responses to the user interaction with the system. The user rating of the conversation responses for fluency was significantly lower than the other metrics since the responses were generated from a scripted group of responses. Users of the system were somewhat unsure of the student's understanding of the concepts when the system would fail on exploratory questions.}}
\label{survey_responses1}
\end{figure*}

\section{THE USER-AI-SUPERVISOR DIALOGUE SYSTEM WITH UNCERTAINTY MODELING}

The \textbf{MDL} and \textbf{UEL} are difficult to meet in CA's for any system and especially for a CA in the context of education. In order to design useful systems, we need to acknowledge that current advances in dialogue systems make having fluent conversations over multiple turns non-trivial. We designed a dialogue system to specifically address user expectation failure and model and data failures. At it's core, the system minimizes bad user experience with uncertainty modeling. When a dialogue system component or a system fails, we redirect control to a supervisor whose task is to bring the conversation back on track. We rely on recent developments in deep learning and uncertainty estimation \citep{gal2016dropout} for each subcomponent of the dialogue system and build an interaction pipeline of user + system + supervisor that can pass the system's control from the dialogue system to the supervisor based on the failure. The supervisor is an application that incorporates an expert user (someone familiar with the scenario and the system's limitations) with the ability to send a response back to the system as a ``Student''. 

\section{Experimental Setting}

% In order to evaluate the teacher question asking strategies we used Instructional Quality Assessment (IQA) \citep{boston2012assessing}.  IQA has been shown to be an effective assessment tool for understanding how teachers ask the different subcategories of questions (see Table \ref{iqa_description}). IQA data collection was performed with two annotators using the labelstudio data labeling platform \citep{LABELSTUDIO}. Our conversational agent comprises of three components (the user of the system, a pre-service or an in-service teacher, the student who is a conversational agent and a supervisor who responds only when the conversational agent redirects control to the user). The user who is the teacher in this context is communicating with the virtual student to teach the concept of scale factor. In traditional classroom scenarios, teachers often use visual aids or a blackboard to explain various mathematical concepts. We have an HTML5 based interactive widget that the teacher can play with to illustrate the concepts to the students (Figure \ref{fig_scale_factor}). 

\subsection{Participant and Model details}
The evaluation included 8 teachers (age between 24-60), each conversing with the system twice. After each session, the users filled up a survey on a 5-point likert scale, as shown in Figure \ref{survey_responses1}.  

The data for the dialogue system was collected through weak supervision based approaches as described in \citep{datta_ws_2021}. A 5-fold cross-validation using the DistilBERT model \citep{sanh2019distilbert} resulted in an \textbf{F1 score: 0.71}. We use ActiveDropout \citep{gal2016dropout} for uncertainty component in the classification task. We used pre-trained BERT \citep{devlin2018bert} sentence representations to train the logistic regression classifier for entity recognition similar to the approach by \citep{nadav2016} (\textbf{precision: 0.84, recall: 0.82, f1: 0.83}). The entity uncertainty (see Figure \ref{fig_uncertainty}) is computed using the softmax probability output on top of the entity extraction task in Table \ref{entity_extraction} (In Appendix).

\subsection{Results}

The system evaluation relied on survey responses from the teachers to judge the effectiveness of the Human-AI collaboration.  The survey \ref{survey_responses1} asked six questions on a 5-point Likert scale. Survey questions that relied on the calibrated uncertainty modules like dialogue act classification and entity recognition (Q1. Did the task get accomplished? Q2. Do you think the student learned anything?) scored higher in user satisfaction than metrics that relied on conversational agent responses (Q4. How fluent was the conversation? Q5. How sensible were the responses?). This is because the system responses were not generative but instead used a template-based approach that seemed repetitive to the user. However, the overall user satisfaction of the system was high since teachers' evaluation of the student involved the student being able to answer the math questions correctly. In this case, the scale factor computation by the conversational agent was deterministic.

\section{Conclusion and Future Work}

One of the main bottlenecks in building conversational agents is the failure mode of deep learning models in the dialogue system's components. Our goal in this paper was to understand the inherent limitations of the dialogue system's different components and effectively use intervention strategies to prevent conversations from derailing. In low resource domains, specifically in education, building new conversational agent scenarios often requires a lot of data, and incorporating uncertainty failure modes can help gather data progressively \citep{mitchell2018never} while maintaining the utility of the tool throughout the process. Uncertainty quantification in deep learning is at it's nascent stage. Despite the current limitations in uncertainty calibration in deep learning we believe future dialogue systems should leverage uncertainty as a core component to improve conversation quality.
\bibliographystyle{plainnat}
\bibliography{sample}

\appendix

\section{Appendix}

\begin{table}[htb]
\caption{\textbf{Entity extraction task for uncertainty modeling}}
\begin{tabular}{llll}
\hline
\textbf{Text}                                     & \textbf{Entities} & \textbf{Relation}    & \textbf{Label} \\ \hline
The length of the object is 5, what is the width? & length, 5, width  & (length, 5)          & True           \\
What is the scale factor?                         & scale factor      & (scale factor, \_\_) & False          \\
No, the length is not 5, the width is.            & length, 5, width  & (width, 5)           & True           \\
No, the length is not 5, the width is.            & length, 5, width  & (length, 5)          & False          \\ \hline
\label{entity_extraction}
\end{tabular}
\end{table}

\end{document}